\documentclass[aps,prl,twocolumn,showpacs,showkeys]{revtex4-1}
\usepackage{graphicx}
\usepackage{amssymb,amsmath}
\usepackage[latin1]{inputenc}
\usepackage{natbib}
\usepackage{epstopdf}
\usepackage{color}
\usepackage{braket}

\begin{document}

\title{Efficient isolation of multiphoton processes and detection of collective states in dilute samples}
\author{Lukas Bruder}
\email{lukas.bruder@physik.uni-freiburg.de}
\author{Marcel Binz}
\author{Frank Stienkemeier}
\affiliation{Physikalisches Institut, Universit\"at Freiburg, Hermann-Herder-Str.\,3, 79104 Freiburg, Germany}

\date{\today}


\begin{abstract}
\noindent A novel technique to sensitively and selectively isolate multiple-quantum coherences in a femtosecond pump-probe setup is presented. Detecting incoherent observables and imparting lock-in amplification, even weak signals of highly dilute samples can be acquired. Applying this method, efficient isolation of one- and two-photon transitions in a rubidium-doped helium droplet beam experiment is demonstrated and collective resonances up to fourth order are observed in a potassium vapor for the first time. Our approach provides new perspectives for coherent experiments in the deep UV and novel multidimensional spectroscopy schemes, in particular when selective detection of particles in dilute gas-phase targets is possible.
\end{abstract}



\maketitle


Multiphoton processes play an important role in many fields of science. 
Light conversion processes such as second harmonic generation or optical parametric amplification are routinely performed in many labs. 
Parametric downconversion\cite{joobeur_spatiotemporal_1994} is the key technique to generate entangled photon pairs used in quantum cryptography applications or to study entanglement properties in various systems\cite{Yuan20101}. 
High harmonic generation has pioneered the development of state-of-the-art coherent light sources in the XUV spectral range having attosecond pulse duration which allow the real-time observation of electron dynamics \cite{Atto_book:2012, Lepine:2014}. 
Likewise, multiphoton absorption in a tight laser focus is used in nonlinear microscopy yielding 3D images of biological tissues with high spatial resolution and great penetration depth\cite{denk_two-photon_1990, lai_nonlinear_2014}.
Furthermore, the energy conversion process in singlet fission incorporates a multiphoton process  and has recently drawn great interest due to its potential application in solar light harvesting\cite{singlet_fission_tetracene:2014, singlet_fission_review:2010}. 

The unique identification and efficient detection of multiphoton processes is however often challenging. 
Monitoring multiphoton absorptions for transitions with small cross sections as a function of intensity is cumbersome and identification by emission spectra is frequently compromised by multistep relaxation pathways or predominant dark relaxation channels. 
Coherent time-resolved spectroscopy offers a different approach to identify multiphoton processes. 
Here, the phase evolution of non-stationary states induced upon optical excitation is monitored. 
This allows not only unambiguous identification of one- and multiphoton processes but also provides high time resolution. 

The phase of superposition states induced by multiphoton transitions - commonly termed multiple-quantum coherences (MQCs) - evolves at a much higher frequency than for one-quantum coherences (1QCs) induced by one-photon transitions. 
In multidimensional spectroscopy, MQC signals have allowed to characterize the influence of electron correlations in molecular systems\cite{kim_measurement_2009}, probing the anharmonicity of molecular potentials\cite{fulmer_pulse_2004} or unraveling the role of high-lying electronic states in ultrafast photoinduced processes\cite{christensson_electronic_2010}.
Furthermore, many-body interactions in a weakly-interacting atomic gas\,\cite{dai_two-dimensional_2012} and in semiconductor nanstructures\,\cite{stone_two-quantum_2009, karaiskaj_two-quantum_2010, turner_coherent_2010} have been revealed by MQC signals. 
The herein probed collective states are only accessible via MQC signals, and in the semiconductor nanostructures binding energies or dephasing times of the observed many-body quasi-particles have been deduced for the first time. 

However, coherences spanning over multiple optical transitions are commonly weak effects. 
Moreover, in many cases higher order signals are masked by intense lower order signals. 
Likewise, monitoring the phase evolution of MQCs demands a high degree of phase stability in the optical excitation scheme. 
Therefore, so far, mainly two-quantum coherence (2QC) signals have been detected\cite{fuller_experimental_2015}. 

In this letter, we present a novel technique to circumvent these issues by performing systematic downsampling of the MQC quantum beats and incorporating lock-in detection for signal enhancement. 
Our method is particular sensitive and very robust, capable of isolating MQC signals of arbitrary high order in a single measurement. 
Furthermore, by applying this technique, we observe for the first time collective resonances up to fourth order in a dilute alkali vapor. Apart from fluorescence detection we demonstrate the sensitivity and universality in the detection scheme by presenting data of mass-selected ions  from a doped helium droplet beam at densities of $\sim 10^{7}$\,cm$^{-3}$. 

\begin{figure}
\centering
  \includegraphics[width=0.96\columnwidth]{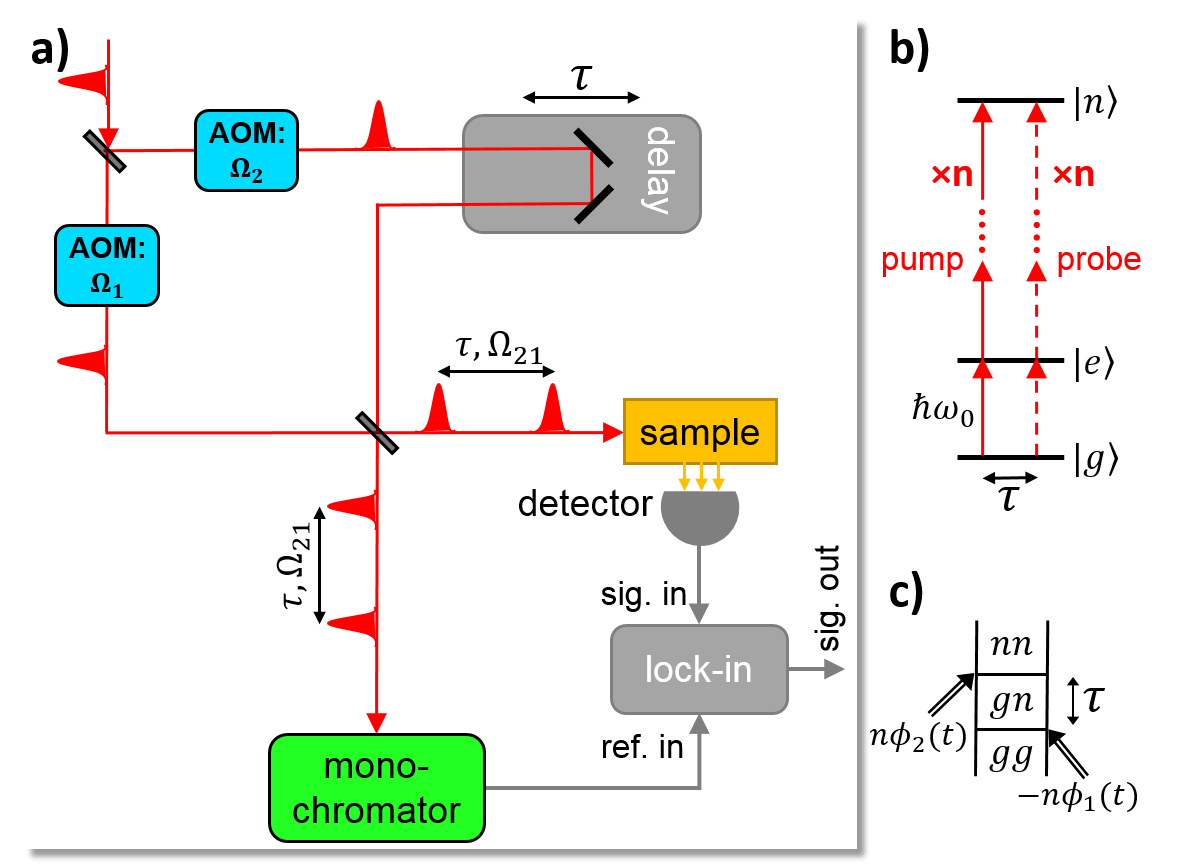}
  \caption{Simplified diagram of the phase-modulated WPI scheme used for the detection of MQCs. The optical setup is shown in (a), diagrammatic WP excitation for a one- and n-photon transition in (b) and a respective Feynman diagram for the n-photon transition in (c).}
  \label{fgr:PM_scheme}
\end{figure}

Our MQC detection scheme is based on electronic wave-packet interferometry (WPI) in a collinear femtosecond (fs) pump-probe scheme (Fig.\,\ref{fgr:PM_scheme}a). 
Briefly - considering at first a one-photon transition as indicated in Fig.\,\ref{fgr:PM_scheme}b  - pump and probe pulses of controllable delay each excite a wave-packet (WP) consisting of a coherent superposition of the initial state $\ket{g}$ and excited state $\ket{e}$. 
Scanning the pump-probe delay $\tau$ results in alternating constructive and destructive interference of these WPs. 
This imparts a typical quantum beat in the recorded pump-probe transient oscillating at the optical transition frequency $\omega_{eg}$ with respect to $\tau$\cite{cina_wave_2002}. 

Such interferometric measurements require a high degree of phase control to obtain a reasonable resolution and signal-to-noise ratio, especially if highly dilute samples are probed as done in our case. 
To reduce these demands, we apply the phase modulation technique established by the Marcus group\cite{tekavec_wave_2006, tekavec_fluorescence-detected_2007}.
In this scheme, an acousto-optic modulator (AOM) is placed in each branch of the interferometer (Fig.\,\ref{fgr:PM_scheme}a), both driven at slightly different frequencies in phase-locked mode. 
The continuous phase sweep between pump and probe pulses imparts an additional modulation in the pump-probe transient, oscillating at the AOM difference frequency, denoted $\Omega_{21}$. 
Thus the signal's phase evolves as $\phi_{sig}(t,\tau)=\Omega_{21}t-\omega_{eg}\tau$\footnote{For simplicity, we consider only the phase of the signals, since it contains all relevant information.}.
Simultaneously, a replica of the modulated pulse train is spectrally filtered in a monochromator yielding $\phi_{ref}(t,\tau)=\Omega_{21}t-\omega_{M}\tau$, where $\omega_M$ denotes the monochromator frequency. 
Referencing this signal to the lock-in amplifier removes the $\Omega_{21}$-modulation in $\phi_{sig}$ and results in a pump-probe transient evolving at a significantly downshifted frequency, yielding for the demodulated signal $\phi_{dm}(\tau)=(\omega_{eg}-\omega_{M})\tau$. 
More details can be found in Refs.\,\cite{bruder_phase-modulated_2015, tekavec_wave_2006, nardin_multidimensional_2013}. 

Likewise, for an n-photon transition ($\ket{g}\rightarrow\ket{n}$, Fig.\,\ref{fgr:PM_scheme}b), the WPI signal is modulated by n-times the AOM difference frequency. 
This relationship has also been observed by Tian and Warren in a two-photon absorption experiment\cite{tian_ultrafast_2002}. 
In the present work, we combine this idea with WPI. 
In the phase-modulated WPI scheme, an n-photon transition yields $\phi_{sig}(t,\tau)=n\Omega_{21}t-\omega_{ng}\tau$ as can be seen from time-dependent perturbation theory. 
An intuitive explanation can also be given in the Feynman diagram representation as commonly used in multidimensional spectroscopy\cite{mukamel1995principles}.
In principle, the acousto-optic phase modulation is equivalent to shot-to-shot phase-cycling\cite{nardin_multidimensional_2013}. 
Thus, $\phi_{sig}$ can be derived from the Feynman diagram shown in Fig.\,\ref{fgr:PM_scheme}c, where  $\phi_j(t)$ (j=1,2) is incremented each laser shot by $\Omega_j T$, with $1/T$ being the laser repetition rate\cite{bruder_phase-modulated_2015}. 

To detect the n-quantum coherence signal, we apply n$^\mathrm{th}$ harmonic lock-in demodulation. 
Inside the lock-in amplifier, the n$^\mathrm{th}$ harmonic of the reference signal is generated, yielding $\phi^{(n)}_{ref}(t,\tau)=n(\Omega_{21}t-\omega_{M}\tau)$ and as output signal $\phi_{dm}(\tau)=(\omega_{ng}-n\omega_M)\tau$. 
In this way, we effectively measure the optical transition frequency in the rotating frame defined by n-times the reference frequency. 
We note that such strong downshifting of optical transition frequencies is not achievable in non-collinear phase-matching where hence much higher phase stability is required. 
Furthermore, the lock-in technique greatly reduces laboratory noise and even weak multiphoton processes are efficiently isolated and amplified. 

Measurements are performed in the same helium droplet molecular beam machine as has been used in previous WPI measurements where interferograms of photoionized desorbed rubidium atoms have been obtained\cite{Mudrich:2008, bruder_phase-modulated_2015}.  Briefly, helium nanodroplets (He$_N$, $N\approx 20\,000$) are doped with single alkali atoms from a heated oven cell. In a differentially pumped detection unit the beam is crossed perpendicularly with fs laser pulses (Ti:Sapphire oscillator, 80\,MHz repetition rate, 200\,fs pulse duration, $\approx 5$\,nJ pulse energy, 40\,$\mu$m focal diameter). The created photoions are mass-filtered in a quadrupole mass spectrometer and its channeltron signal is fed into lock-in amplifieres. 
Fluorescence detection has been performed in a heated alkali vapor cell by means of a photodiode, impedance converted by a home-build preamplifier. 



\begin{figure}
	\centering
	\includegraphics[width=0.7\columnwidth]{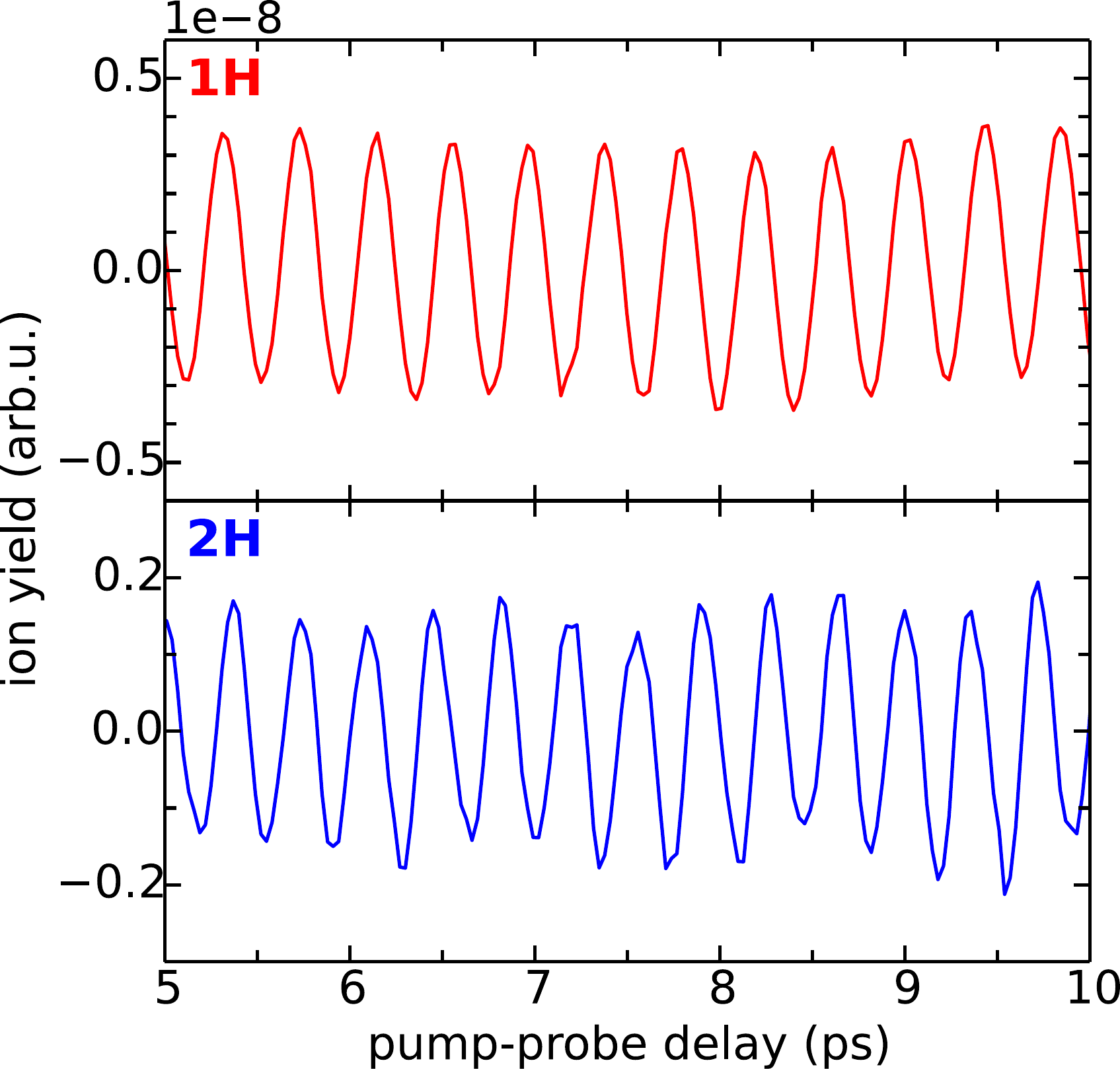}
	\caption{Interferograms recorded in a single measurement referencing to the 1$^{\rm st}$ and 2$^{\rm nd}$ harmonic of $\Omega_{21}$. The downshifting by $\omega_{M}$ and $2\omega_{M}$, respectively, results in similar oscillation periods with respect to the pump-probe delay $\tau$.}
	\label{fgr:Interferograms_Rb}
\end{figure}
\begin{figure}
	\centering
	\includegraphics[width=0.95\columnwidth]{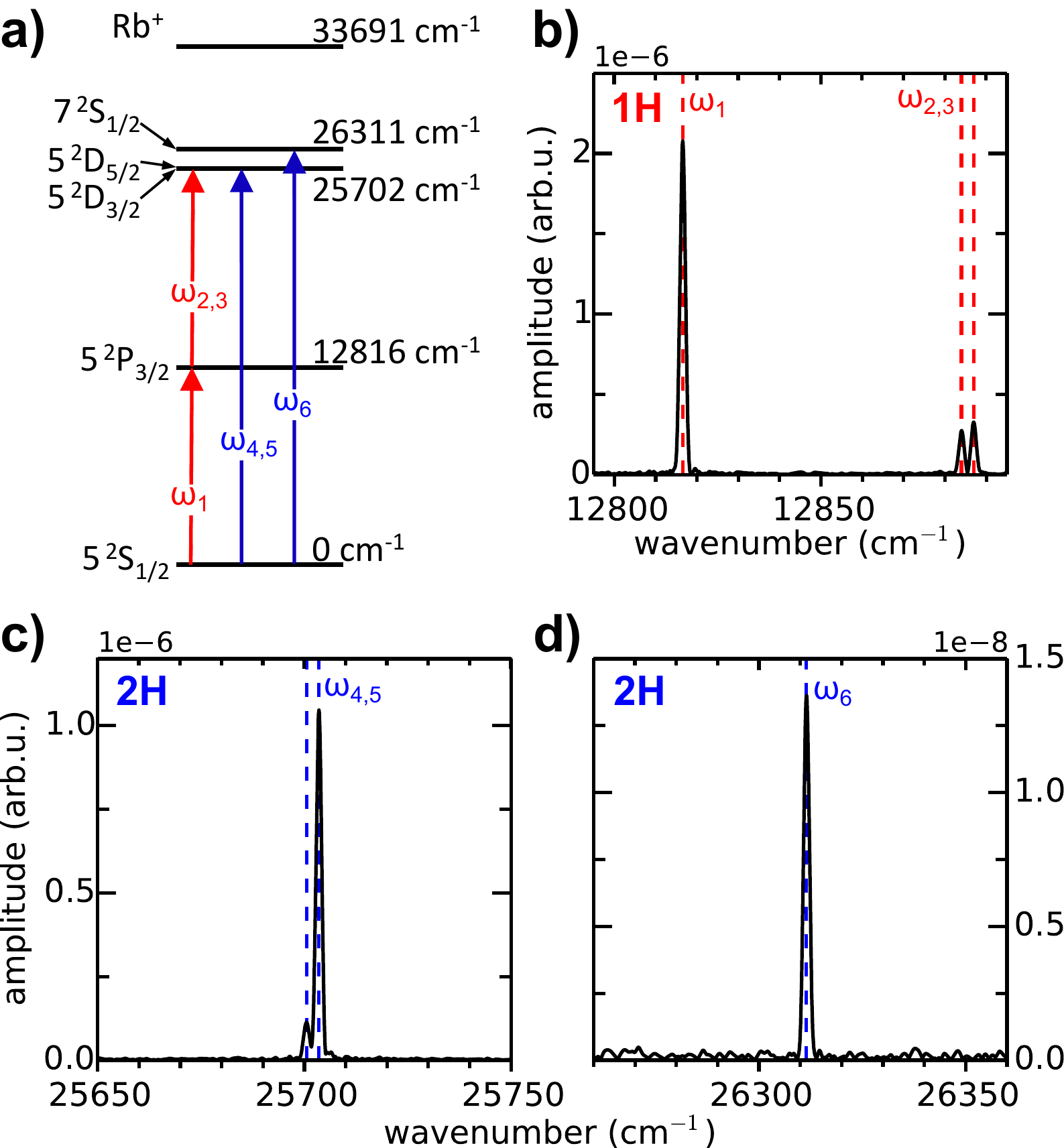}
	\caption{Level diagram of the involved Rb electronic states (a). Fourier transformed spectra of the measured phase-modulated interferograms referenced to $\omega_{M}$ (b) and $2\omega_{M}$ (c),(d). Dashed vertical lines show tabulated transitions\cite{NIST_ASD}.}
	\label{fgr:MQC_Rb}
\end{figure}

As a first demonstration of the method we record phase-modulated interferograms of the mass-selected photoion yield of rubidium (Fig.\ \ref{fgr:Interferograms_Rb}), referenced to the 1$^{\rm st}$ and 2$^{\rm nd}$ harmonic of $\phi_{ref}=\Omega_{21}t-\omega_M \tau$. 
The laser was set to 12904\,cm$^{-1}$, enabling the resonant excitation of both the 5$^2$P$_{3/2}$ and the 5$^2$D$_{3/2, 5/2}$ states (cf. Fig.\ \ref{fgr:MQC_Rb}a), and subsequent ionization with 3 photons in total. The Fourier transformed spectra (Fig.\ \ref{fgr:MQC_Rb}b-d) selectively yield the 1-photon transitions (denoted $\omega_{1,2,3}$, Fig.\ \ref{fgr:MQC_Rb}b) or the two-photon transitions (denoted $\omega_{4,5,6}$, Fig.\ \ref{fgr:MQC_Rb}c+d). In order to include the nonresonant excitation to the 7$^2$S$_{1/2}$ state ($\omega_6$), the laser central wavenumber was shifted to 13182\,cm$^{-1}$. 
We scan the pump-probe delay up to 40\,ps to obtain well resolved fine structure components of the 5D state. 
In the two measurements, the monochromator was set to 12898 and 13134\,cm$^{-1}$, respectively, leading to downshifted oscillations $< 3$\,THz for both first and second harmonic detection. 
Note the excellent signal-to-noise ratio of mass-selected ions at target densities of $\approx 10^{7}$\,cm$^{-3}$. 
The same signals are observed with fluorescence detection in a vapor cell; however, 2-photon signals are augmented by the enhanced photoionization of the higher-lying states.

\begin{figure}
\centering
  \includegraphics[width=0.9\columnwidth]{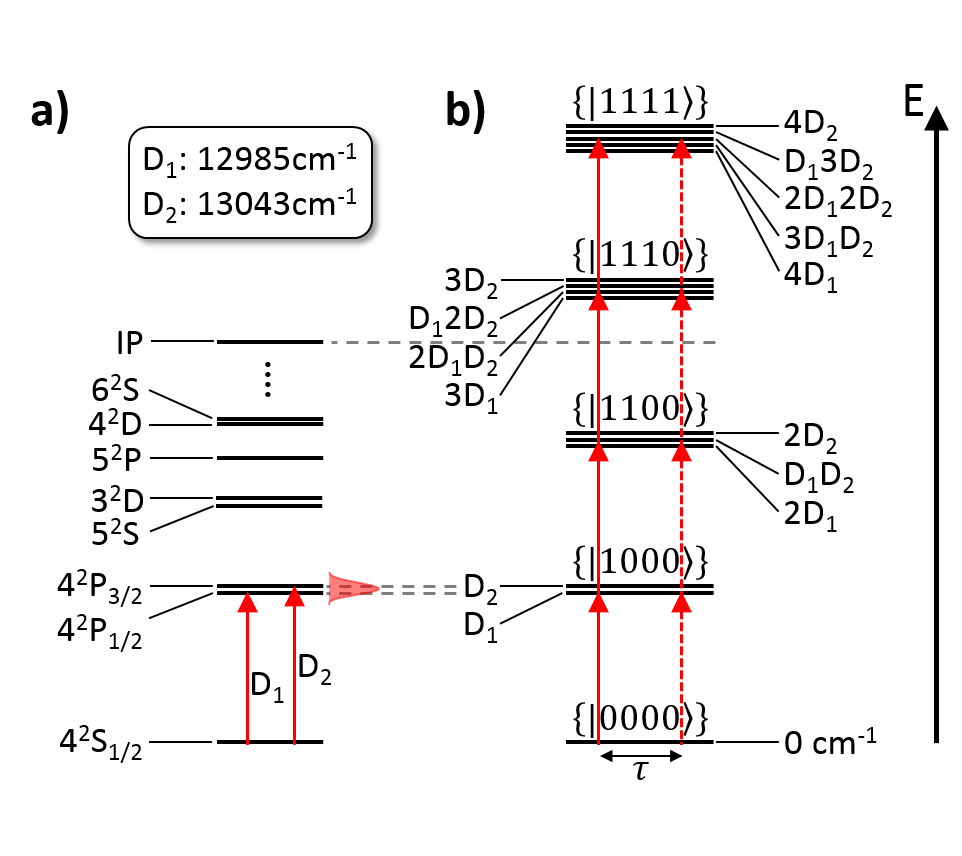}
  \caption{Level diagram for atomic potassium and wavenumbers for the excited D$_1$ and D$_2$ lines in (a), and corresponding collective energy states for an ensemble of four potassium atoms described in a product basis in (b). For clearer visualization in (a), states lying between the 6$^2$S and the ionic potential (IP) are not shown. Since the laser is only resonant to the D1 and D2 line transitions, only these excitations are considered for the collective energy levels.}
  \label{fgr:CR_levels}
\end{figure}
In a second set of measurements, we employ the established MQC detection technique to reveal collective resonances in a potassium vapor of 10$^{10}$\,cm$^{-3}$ density using fluorescence detection. 
D$_1$ and D$_2$ lines are excited simultaneously with the unfocused pump-probe beam (1.5\,mm diameter) tuned to 13002\,cm$^{-1}$. 
Due to the limited laser bandwidth (89\,cm$^{-1}$ FWHM), multiphoton transitions to higher-lying states are not present and the potassium atom reduces to an effective three-level system (Fig.\,\ref{fgr:CR_levels}a). 
However, if considering an ensemble of atoms described in a product basis, one can assign new collective energy states, yielding a ladder-type energy structure as exemplary shown for an ensemble of four atoms in Fig.\,\ref{fgr:CR_levels}b. 
Such collective states may be assigned to any many-body system independent whether significant interactions among the constituents are present or not. 
In our experiment, we estimate the dipole-dipole interaction of the D line excitations being below $10^{-7}\,$cm$^{-1}$ and thus being negligible. 
Hence, in the presented coherent pump-probe experiment, the atoms are primarily coupled via the intense fs laser pulses. 
The simultaneous stimulation of atoms initiates a collective oscillation of the optically induced non-stationary superposition states. 
We note, that these collective states are not accessible by direct excitation in the visible to ultraviolet spectral range. 
Likewise, photons in the near infrared spectrum are emitted by the sample. 

\begin{figure}
	\centering
	\includegraphics[width=0.99\columnwidth]{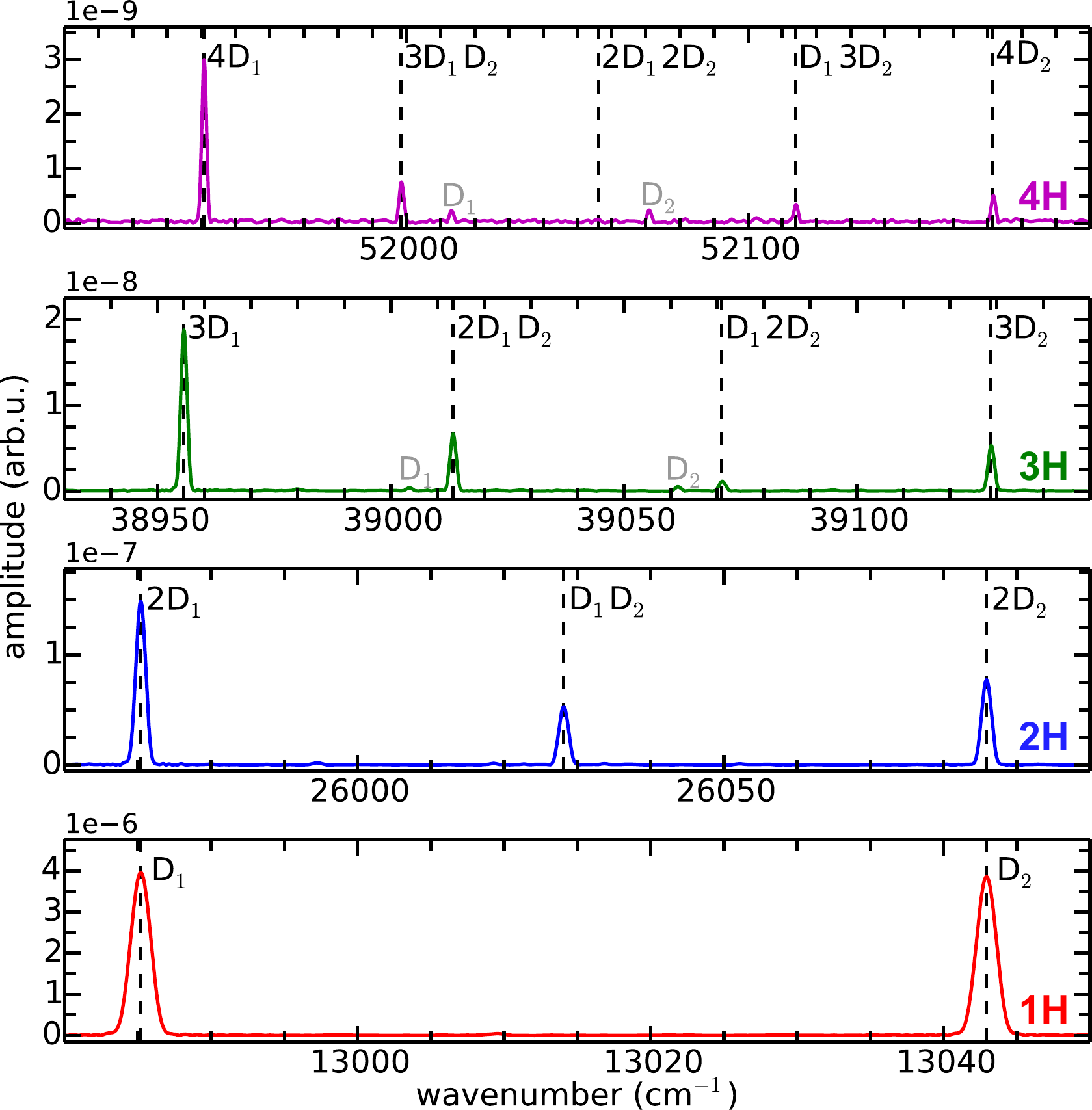}
	\caption{Collective resonances in potassium vapor for 1$^{\rm st}$-4$^{\rm th}$ harmonic lock-in demodulation (labeled 1H-4H). Black dashed lines indicate the predicted energies of the collective states. Their labels correspond to the notation introduced in Fig.\ref{fgr:CR_levels}. Peaks with gray labels correspond to leak signals from the 1$^{\rm st}$ harmonic.}
	\label{fgr:MQC_K}
\end{figure} 
We record MQC quantum beats for 1$^{\rm st}$ -- 4$^{\rm th}$ harmonic demodulation in a single measurement with four lock-in amplifiers connected in parallel. 
The corresponding Fourier spectra reveal resonances precisely at the predicted collective energies (Fig.\,\ref{fgr:MQC_K}). 
With each harmonic, the signal decreases by about one order of magnitude. 
Due to the undersampling effect, demands on phase stability are in all harmonics of the same order, resulting in an excellent signal-to-noise ratio, which allows acquisition of even higher-order collective states. 
The plotted highest quantum coherence (4D$_2$ state) corresponds to a transition in the deep ultraviolet spectral range (192\,nm). 
To the author's knowledge, electronic WPI or other coherent time-resolved spectroscopies involving transitions of such high energy have only been successfully conducted for collective states in semiconductors, where a much more complicated setup has been used\cite{turner_coherent_2010}; thus emphasizing the potential in the simplicity of the presented method.

While peak positions fit perfectly with prediction, peak amplitudes are more difficult to calculate. 
On the one hand, excitation paths leading to cross peaks (e.g. D$_1$D$_2$ peak in subpanel 2H, Fig.\ \ref{fgr:MQC_K}) partially interfere destructively resulting in signal reduction. 
On the other hand, our observations indicate that peak heights are compromised by saturation effects, which occur for D line excitations in alkali atoms at very low laser intensities. 
A detailed analysis of amplitudes as well as vapor density dependencies will be published elsewhere. 

As visible in subpanels 3H and 4H of Fig.\,\ref{fgr:MQC_K}, leak signals from one-photon transitions may also occur. 
These signals arise from spurious 1$^{\rm st}$ harmonic signal components in the lock-in referencing. 
Since the appearance of leak signals strongly differs when employing devices from different manufacturers, we assign this to a technical artifact. 
The leak signals are unambiguously identified even in unknown systems by their dependence on the monochromator frequency. 
Other sources for artifacts in the experiment could be frequency upconversion effects among the spontaneous emitted photons, nonlinear processes in the photo detector and saturation of amplifiers in the electronic circuits. 
However, a nonlinear behavior in all mentioned cases has been excluded by systematic investigations. 
Artifacts due to the laser repetition rate being greater than the decay rate of excited states are excluded since we observe the collective resonances also when employing a 5\,kHz repetition rate laser.

We detected analog many-body resonances in the photoion yield of the Rb doped droplet measurements as well where densities are even orders of magnitudes smaller (not shown). 
Further, for $\geq 3$-body correlations we have observed a signal depletion for increasing density, indicating a collective behavior beyond the coupling induced by the external laser field; these results will be published in a separate paper. 


In conclusion, we introduced a new scheme to selectively and sensitively record multiple-quantum coherences in dilute alkali vapors and supersonic alkali-doped helium droplet beams, demonstrated by spectra up to the 4$^{\rm th}$ order. 
So far, MQCs have been almost exclusively recorded only up to 2nd order in four- or higher-wave mixing experiments which are not feasible at dilute gas-phase targets. 
Our two-pulse approach is thus much simpler and principly allows detection of arbitrary high orders. 
If desired, the technique can be integrated in any kind of multidimensional spectroscopy scheme and will also work with shorter pulse lengths. 
In this way the presented method provides new perspectives to design higher order multidimensional spectroscopic schemes. 
It allows simultaneous isolation of multiple processes thus providing a multilayer picture of the investigated system within a single measurement. 
The incoherent probing allows for highly selective observables; here demonstrated for mass-selected ions. 
However, energy or angular-resolved electrons, or any action or depletion spectroscopic signal can be used. 
In addition, the lock-in amplification provides an extremely sensitive method suitable for high order photon processes or single-molecule studies.

 Stimmulating discussions with Alexander Eisfeld, Frank Schlawin and Manuel Gessner are gratefully acknowledged.  The project is supported by the Deutsche Forschungsgemeinschaft within the IRTG 2079. L$.$ B$.$ thanks the Evangelisches Studienwerk e$.$V$.$ Villigst for financial support.

\end{document}